\documentclass[11pt]{article}
\usepackage{graphicx}

\newcommand{\BaBarYear}       {01}
\newcommand{\BaBarNumber}     {01}
\newcommand{\SLACPubNumber} {8924}

\newcommand{\BaBarType}     {CONF}  

\input pubboard/babarsym

\newcommand{\bztoccz}{$B^0\rightarrow D^{(*)-}D^{(*)+}K^0_S$}
\newcommand{\bdstdzk}{$B^0\rightarrow D^{*-}D^{0}K^+$}
\newcommand{\bdstdstzk}{$B^0\rightarrow D^{*-}D^{*0}K^+$}

\newcommand{\btoddk}{\ensuremath{B\ra D^{(*)} D^{(*)} K}}
\newcommand{\btodsdsk}{\ensuremath{B^+ \to D^{*-} D^{*+} K^+}}
\newcommand{\de}{\ensuremath{\Delta E}}
\def\Dpm     {\ensuremath{D^\pm}}

\long\def\inst#1{\par\nobreak\kern 4pt\nobreak
{\it #1}\par\vskip 10pt plus 3pt minus 3pt}

\begin{document}
{\pagestyle{empty}

\begin{flushright}
\babar-\BaBarType-\BaBarYear/\BaBarNumber \\
SLAC-PUB-\SLACPubNumber \\
July, 2001 \\
\end{flushright}

\par\vskip 3cm

\begin{center}
\Large \bf \boldmath Investigation of $B \rightarrow D^{(*)}\overline D^{(*)}K$ decays with the \babar\ 
detector
\end{center}
\bigskip

\begin{center}
\large The \babar\ Collaboration\\
\mbox{ }\\
July 20, 2001
\end{center}
\bigskip \bigskip

\begin{center}
\large \bf Abstract
\end{center}
Using about 23M $B \overline B$ events collected in 1999-2000 with the \babar\ 
detector, we report the observation of several hundred 
$B \rightarrow D^{(*)}\overline D^{(*)}K$ decays 
with two completely reconstructed $D$ mesons. The preliminary branching fractions of the 
low background decay modes $B^0\rightarrow D^{*-}D^{(*)0}K^+$ are determined to be 
$ {\cal B}(B^0 \rightarrow  D^{*-}D^{0}K^+) = (2.8 \pm 0.7 \pm 0.5)\times 10^{-3} $ and 
$ {\cal B}(B^0 \rightarrow D^{*-}D^{*0}K^+) = (6.8 \pm 1.7 \pm 1.7)\times 10^{-3} $. 
Observation of a significant number of candidates in the color-suppressed 
decay mode $B^+\rightarrow D^{*+}D^{*-}K^+$ is reported with a preliminary branching fraction 
$ {\cal B}(B^+\rightarrow D^{*+}D^{*-}K^+)= (3.4\pm 1.6\pm 0.9)\times 10^{-3}.$

\vfill
\begin{center}
Submitted to the\\ 20$^{th}$ International Symposium 
on
Lepton and Photon Interactions at High Energies, \\
7/23---7/28/2001, Rome, Italy
\end{center}

\vspace{1.0cm}
\begin{center}
{\em Stanford Linear Accelerator Center, Stanford University, 
Stanford, CA 94309} \\ \vspace{0.1cm}\hrule\vspace{0.1cm}
Work supported in part by Department of Energy contract DE-AC03-76SF00515.
\end{center}
}

\newpage

\begin{center}
\small

The \babar\ Collaboration,
\bigskip

B.~Aubert,
D.~Boutigny,
J.-M.~Gaillard,
A.~Hicheur,
Y.~Karyotakis,
J.~P.~Lees,
P.~Robbe,
V.~Tisserand
\inst{Laboratoire de Physique des Particules, F-74941 Annecy-le-Vieux, France }
A.~Palano
\inst{Universit\`a di Bari, Dipartimento di Fisica and INFN, I-70126 Bari, Italy }
G.~P.~Chen,
J.~C.~Chen,
N.~D.~Qi,
G.~Rong,
P.~Wang,
Y.~S.~Zhu
\inst{Institute of High Energy Physics, Beijing 100039, China }
% ...

\end{center}\newpage

\setcounter{footnote}{0}

\section{Introduction}
\label{sec:Introduction}
Decays of $B$ mesons that include a charmed and an anti-charmed meson
 are expected to occur through the $b$ to $c$ quark transitions 
$\overline b \to \overline c W^+$, where the $W^+$ materializes as a $c \overline s$ pair. 
These transitions are responsible for most of the $D_s$ production in $B$ 
decays. $D_s$ production has been thoroughly 
studied in experiments running at the $\Upsilon(4S)$ resonance
\cite{ref:dsdargus,ref:dsdcleo,ref:dsdcleo2}. The inclusive rate for 
 $D_s$ production in $B$ decays was recently measured by \babar, where 
a preliminary branching fraction  \cite{ref:dsdbabar}:
$$ {\cal B} (B \rightarrow D_s X) = {\mathrm (10.93 \pm 
0.19_{stat} \pm 0.58_{syst}\pm 2.73_{\phi \pi})}  \% $$
is reported.
\par Until 1994, it was believed that the $c \overline s$ quarks
 would hadronize dominantly as $D_s^{+(*)}$ mesons. Therefore, 
the  branching fraction $b \to c \overline c s$ 
was computed from the inclusive $B \to D_s \, X $, 
$B \to (c \overline c)\, X $ and 
$B \to \Xi_c \, X $ branching fractions, leading 
 to ${\cal B}(b \rightarrow c \overline c s)={\mathrm 15.8 \pm 2.8 \%}$ 
\cite{ref:browder2}.
Theoretical calculations are unable to simultaneously describe this low 
branching fraction and the semileptonic branching fraction of the $B$ 
meson \cite{ref:bigi}. It has been conjectured 
\cite{ref:buchalla} that ${\cal B}(b \to c \overline c s)$ is in fact 
larger and that decays $B \to D \overline D K\,(X)$ 
(where D can be either a $D^0$ or a \Dpm)   
 could contribute significantly. This might also include possible decays to  
orbitally-excited $D_s$ mesons, 
$B \to \overline D^{(*)} D_s^{**}$, 
followed by $D_s^{**}\to D^{(*)}\, K$. 

Some experimental support for this picture has been published in the last 
few years. The most significant results are the evidence for wrong-sign $D$ 
production in $B$ decays (CLEO), yielding 
${\cal B}(B \to D\, X)=7.9\pm 2.2\% $ \cite{ref:cleoupv}, and 
the observation of a small number of completely reconstructed 
\btoddk\  decays, by both CLEO \cite{ref:cleoddk} and ALEPH \cite{ref:alephddk}.
\par
\begin{figure}[t]
\begin{center}
\scalebox{1.}{
\includegraphics[height=6.5cm]{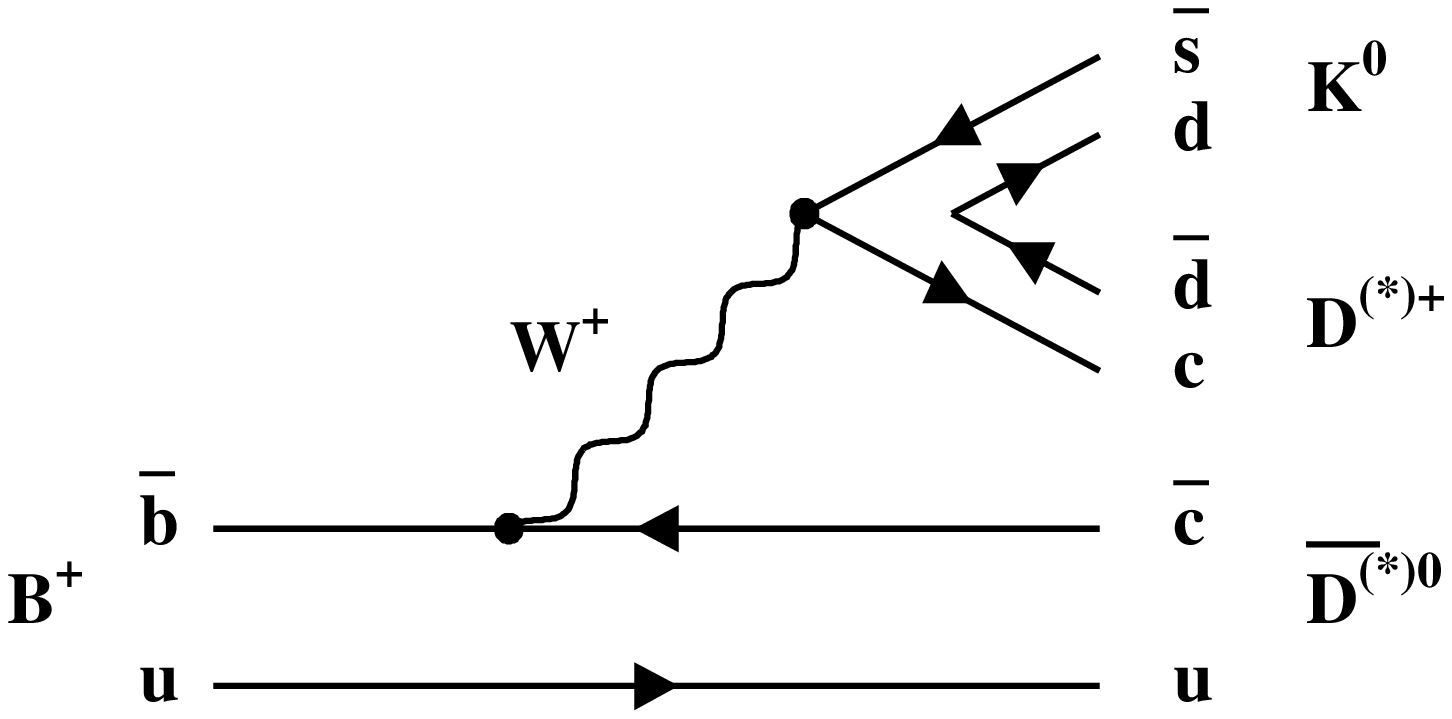}
\includegraphics[height=6.5cm]{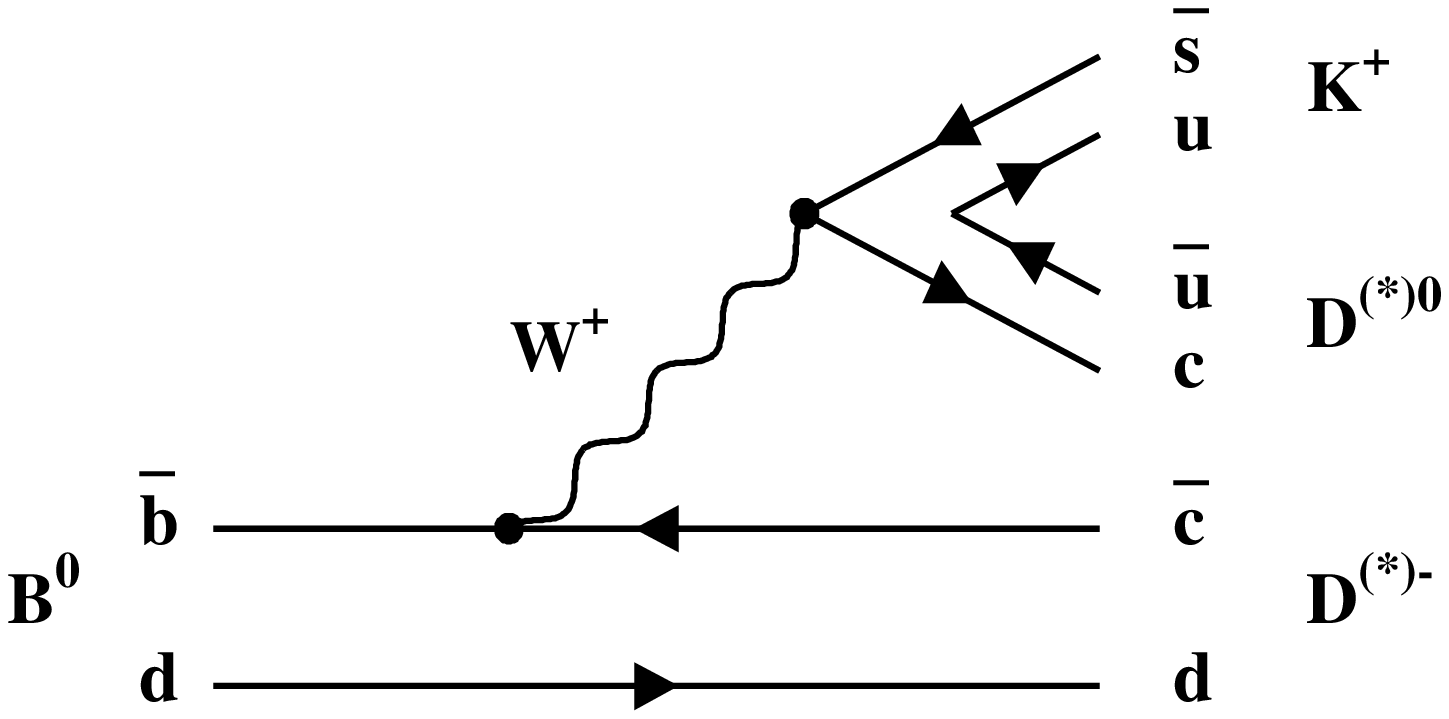}}
\scalebox{1.}{
\includegraphics[height=6.5cm]{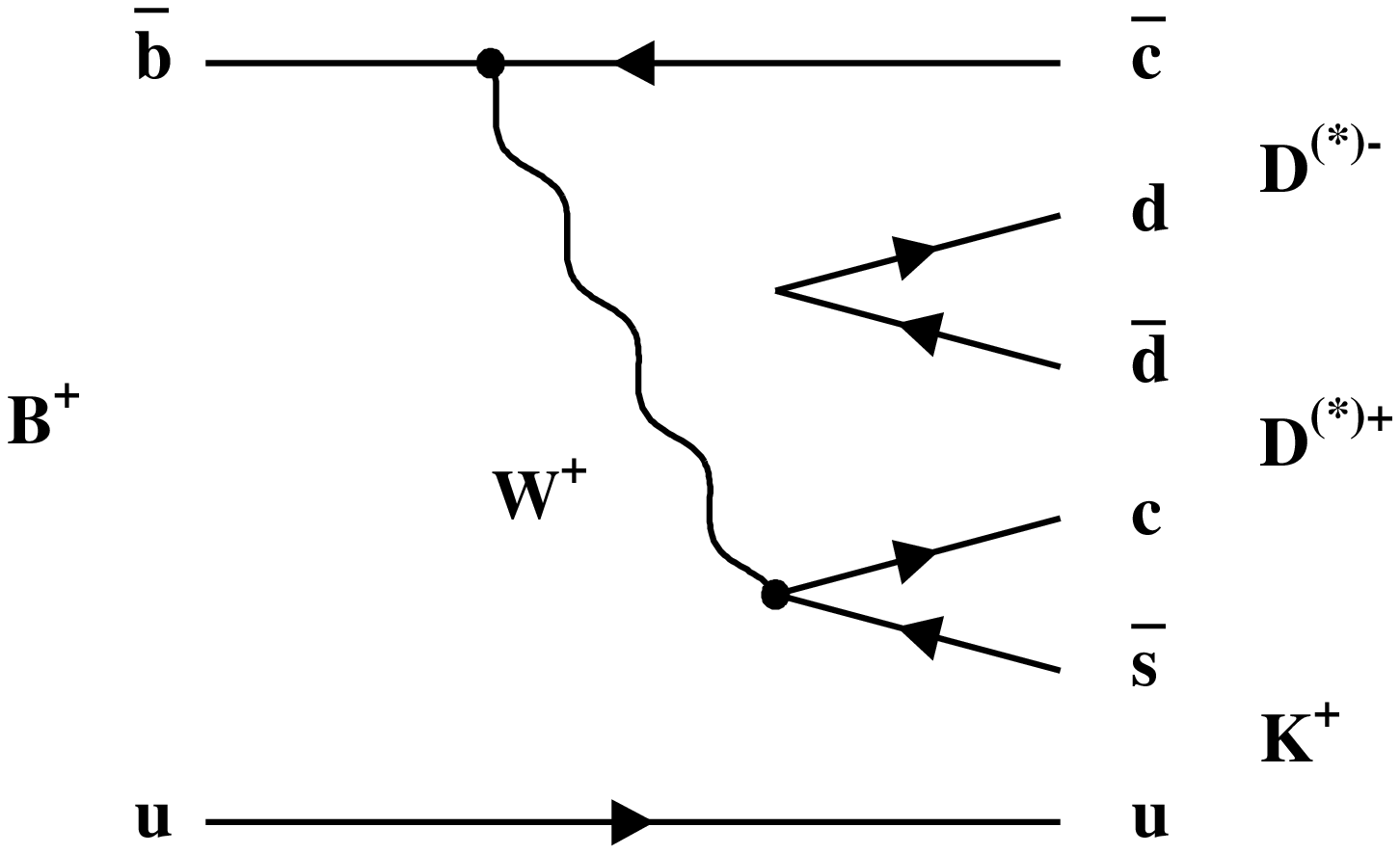}
\includegraphics[height=6.5cm]{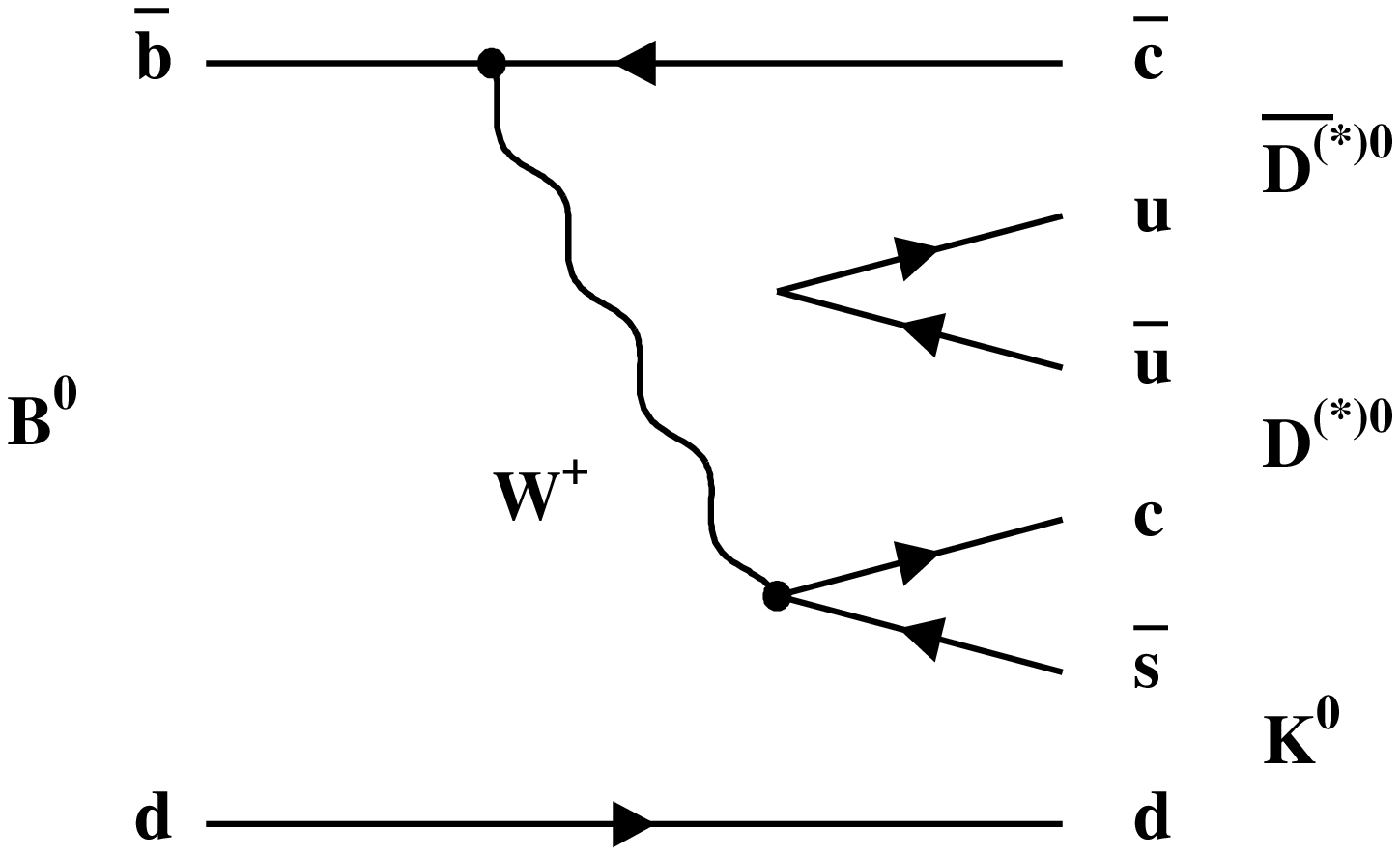}}
\scalebox{1.}{\includegraphics[height=6.5cm]{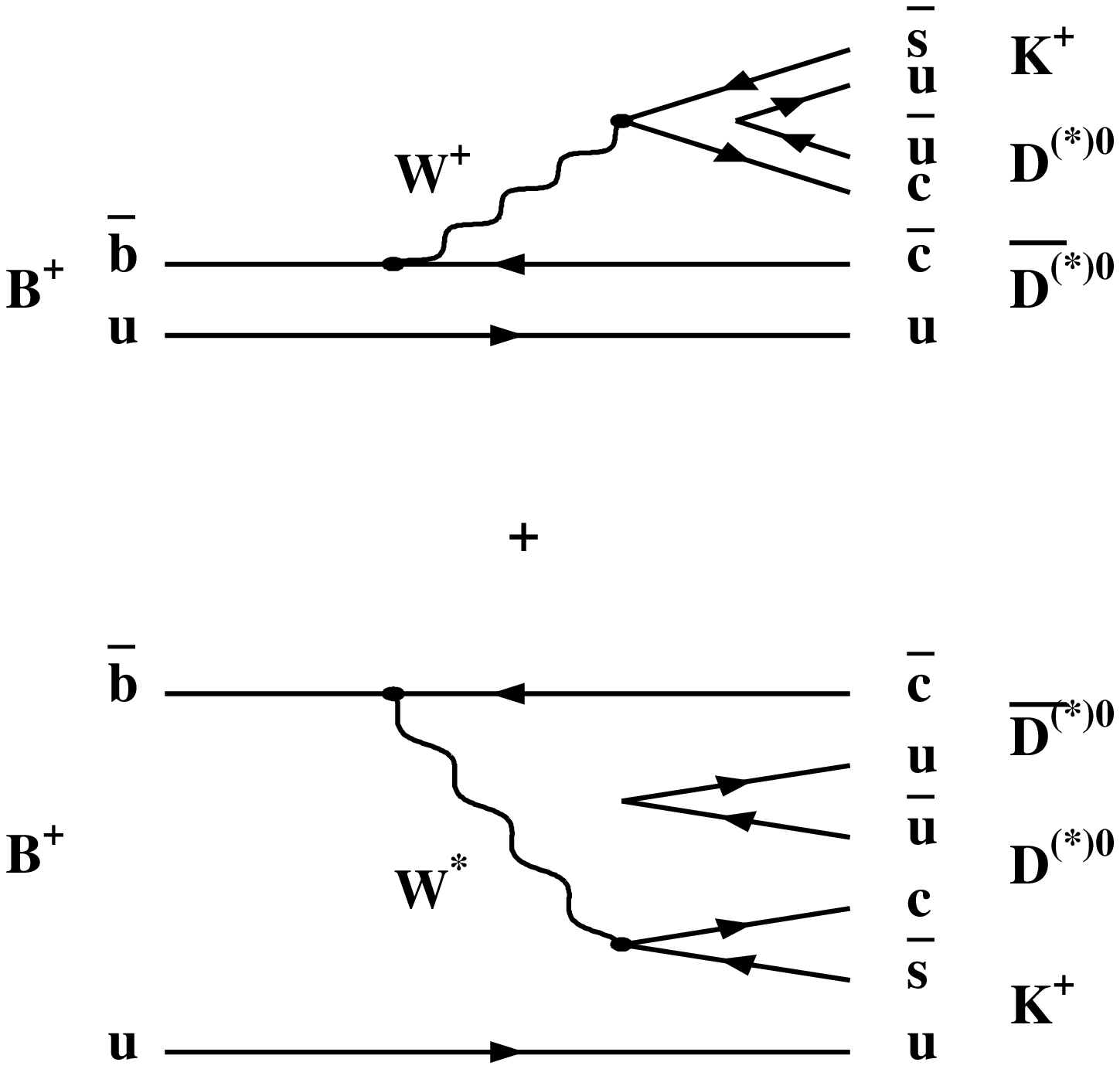}
\includegraphics[height=6.5cm]{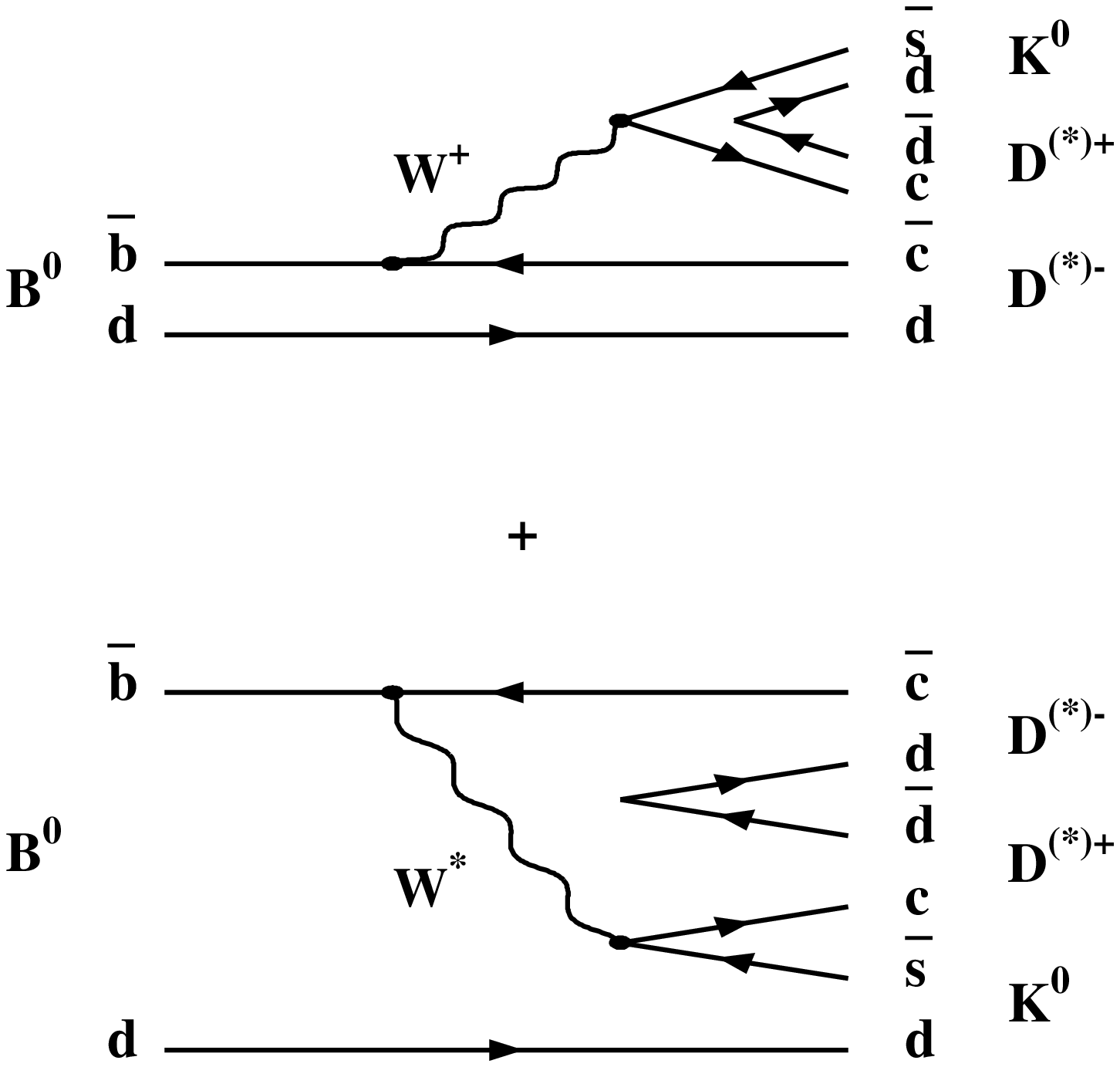}}
\caption{$DDK$ decays proceed through external only diagrams (top), internal only diagrams (2nd line) and both (last lines)}
\label{Fi:external}
\end{center}
\end{figure}

 \btoddk\ decays can occur through three different processes: 
pure external diagrams, pure internal (color-suppressed) diagrams 
 and the sum of both. Fig. \ref{Fi:external} shows the three possible types
of decays for charged and neutral $B$'s.

\par In \babar, the high statistics available allow comprehensive 
investigations to be made of the $b \to c \overline c s$ transitions. In the 
analysis described in this paper, we present evidence for the decays  
$B\to D^{(*)} \overline D^{(*)} \KS$  and $B\rightarrow D^{(*)} \overline D^{(*)} K^\pm$, 
using events in which both $D$'s are completely reconstructed. 
After describing the data sample and the event 
selection, we show the $D^{(*)}\overline D^{(*)}K$ signals  for 
the sum of all $B$ submodes. The branching fractions for some of the cleanest 
modes, such as $B^0 \rightarrow D^{*-} D^{(*)0} K^+$, are computed and the main 
systematic errors are discussed. Observation of several candidates in the 
color-suppressed decay mode  \btodsdsk\ is also reported.  

\section{\boldmath The \babar\ detector and dataset}
\label{sec:babar}
The study reported here uses 20.7\invfb\ of data collected at the 
$\Upsilon(4S)$ resonance  with the \babar\ detector, corresponding to 
 $(22.7 \pm 0.4)\times 10^6$ $B \overline B$ pairs. 
\par The \babar\ detector is a large-acceptance solenoidal 
spectrometer (1.5 T) described in detail elsewhere~\cite{ref:babar}. 
The analysis described below makes use of charged track and 
$\pi^0$ reconstruction and charged particle identification.
Charged particle trajectories are measured by a 5 layer double-sided 
silicon vertex tracker (SVT) and a 40-layer drift chamber (DCH), which also 
provide ionisation measurements ($dE/dx$) used for particle 
identification. Photons and electrons are measured in the electromagnetic 
calorimeter (EMC), made of 6580 thallium-doped CsI crystals constructed 
in a non-projective barrel and forward endcap geometry. Charged $K/\pi$ 
separation up to 4\gevc\ in momentum is provided by a detector of internally 
reflected Cherenkov light (DIRC), consisting of 12 sectors of quartz 
bars that carry the Cherenkov light to an expansion volume filled with water
and equipped with 10752 photomultipliers. 

\section{\boldmath Analysis strategy}
\label{sec:Analysis}
The $B^0$ and $B^+$ mesons\footnote {Charge-conjugate reactions are implied 
throughout this note.} are reconstructed in a sample of multihadron events 
for all possible $D \overline D K$ modes, namely 
$B^0\rightarrow D^{(*)-}D^{(*)0}K^+$, 
$D^{(*)-}D^{(*)+}K^0$, $\overline D^{(*)0}D^{(*)0}K^0$ and 
$B^+\rightarrow \overline D^{(*)0}D^{(*)+}K^0$, 
$\overline D^{(*)0}D^{(*)0}K^+$, $\overline D^{(*)+}D^{(*)-}K^+$. 
 \par The \KS candidates are reconstructed from two oppositely charged tracks
 coming from a common vertex displaced from the interaction point by at least 
0.2 cm and having an invariant mass within $\pm 9 \mevcc$ of the nominal $K^0$ 
mass. The \piz 
candidates are reconstructed from pairs of photons, each with an energy 
greater than 30\mev, which are required to have a mass 
$115<M_{\gamma \gamma}<150\mevcc$. The \piz from \Dstarz must 
have a momentum $70<p^*(\gamma\gamma)<450\mevc$ in the $\Upsilon(4S)$ frame, 
while the \piz from $D^0\rightarrow K^-\pi^+\pi^0$ must have an energy $E(\piz)>200\mev$. 
Finally, a mass-constraint fit is applied to all the 
\piz candidates to improve the energy resolution.
\par The $D^0$ and $D^+$ mesons are reconstructed in the modes 
$D^0\rightarrow K^-\pi^+$, $K^-\pi^+\pi^0$, $K^-\pi^+\pi^-\pi^+$ and 
$D^+\rightarrow K^-\pi^+\pi^+$, by selecting track
combinations within $\pm 2\sigma$ or $\pm 3\sigma$ of the nominal $D$ mass, 
where $\sigma$ is the mass resolution for the $D$ decay mode considered and 
the tighter $2\sigma$ mass interval is applied for $B$ modes with a larger 
combinatorial background. The $K$ and $\pi$ tracks are required to be well 
reconstructed in the tracking detectors and to originate from a common vertex. 
Charged kaon identification, 
with information from the Cherenkov angle in the DIRC and from $dE/dx$ 
measurements in the drift chamber and in the vertex detector, is required 
for most $D$ decay modes, as well as for the \Kpm from $B$'s.
\par $D^*$ candidates are reconstructed in the modes 
$D^{*+}\rightarrow D^0\pi^+$, $D^{*0}\rightarrow D^0\pi^0$ and    
$D^{*0}\rightarrow D^0\gamma$, by combining a $D^0$ candidate with a $\pim$, $\piz$, or photon.
A $\pm 3 \sigma$ interval around the nominal $\Delta M= M(D^*)-M(D^0)$ mass difference is used
to select $D^*$'s. Partial reconstruction of $D^{*0}$'s (no \piz or $\gamma$ 
reconstruction) is also used in the \bdstdstzk\ mode, as explained below.
\par $B$ candidates are reconstructed from the $D^{(*)}$, $\overline D^{(*)}$ 
and $K$ candidates. A mass constraint is applied to all the intermediate 
particles ($D^0$, $D^-$, \KS). Since the $B$ mesons are produced via \epem 
$\rightarrow$ \upsbb, the energy of the $B$ in the $\Upsilon{( 4S)}$ frame 
is given by the beam energy $E_{beam}^*$, which is measured much more 
precisely than the energy of the $B$ candidate. Therefore, to isolate the 
$B$ meson signal, we use two kinematic variables: \de, the difference 
between the reconstructed energy of the $B$ candidate and the beam energy in 
the center of mass frame, and \mes, the beam energy substituted mass, 
defined as $$\mes = \sqrt{E_{beam}^{*2}-p_B^{*2}}$$
where $p_B^*$ is the momentum of the reconstructed $B$ in the $\Upsilon{( 4S)}$ 
frame. Signal events will have \de\ close to 0 and \mes close to the $B$ meson 
mass, 5.729\gevcc. When several candidates are selected per event in a 
specific $B$ submode ({\it e.g.} $B^+\rightarrow D^0\overline D^0 K^+$), a $\chi^2$ 
value, taking into account the difference between the measured and the PDG 
values of the $D$ masses and of the $\Delta M$ (for $D^*$'s) is constructed and 
only the candidate with the lowest $\chi^2$ value is kept for the given 
submode.  
\section{Evidence for signal in the sum of all $B$ submodes}

We present here the distributions obtained by summing all possible
\btoddk\ decay channels, for neutral and charged $B$ decays respectively.

\par
\begin{figure}[H]
\begin{center}
\scalebox{.35}{
\includegraphics{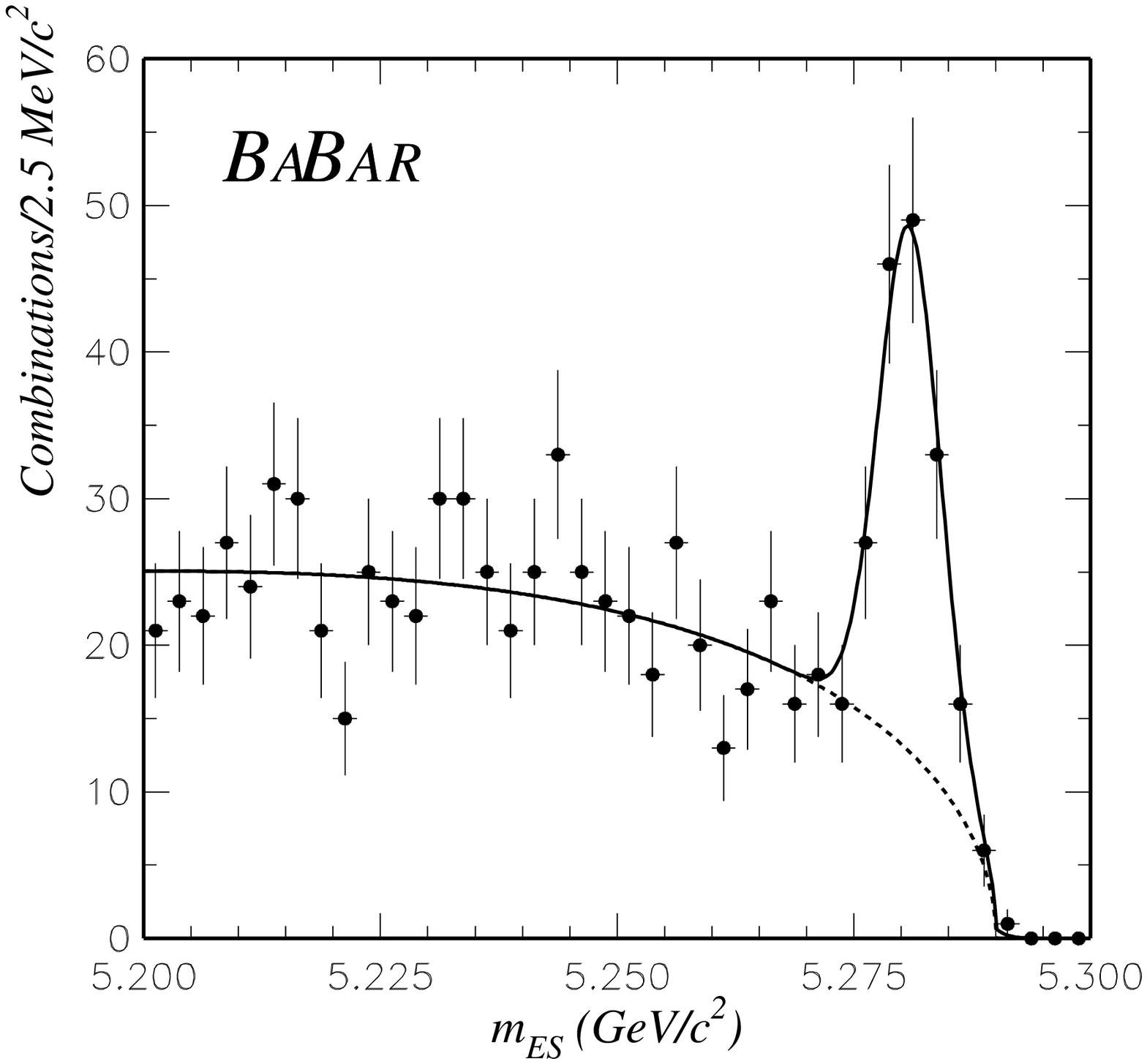}
\includegraphics{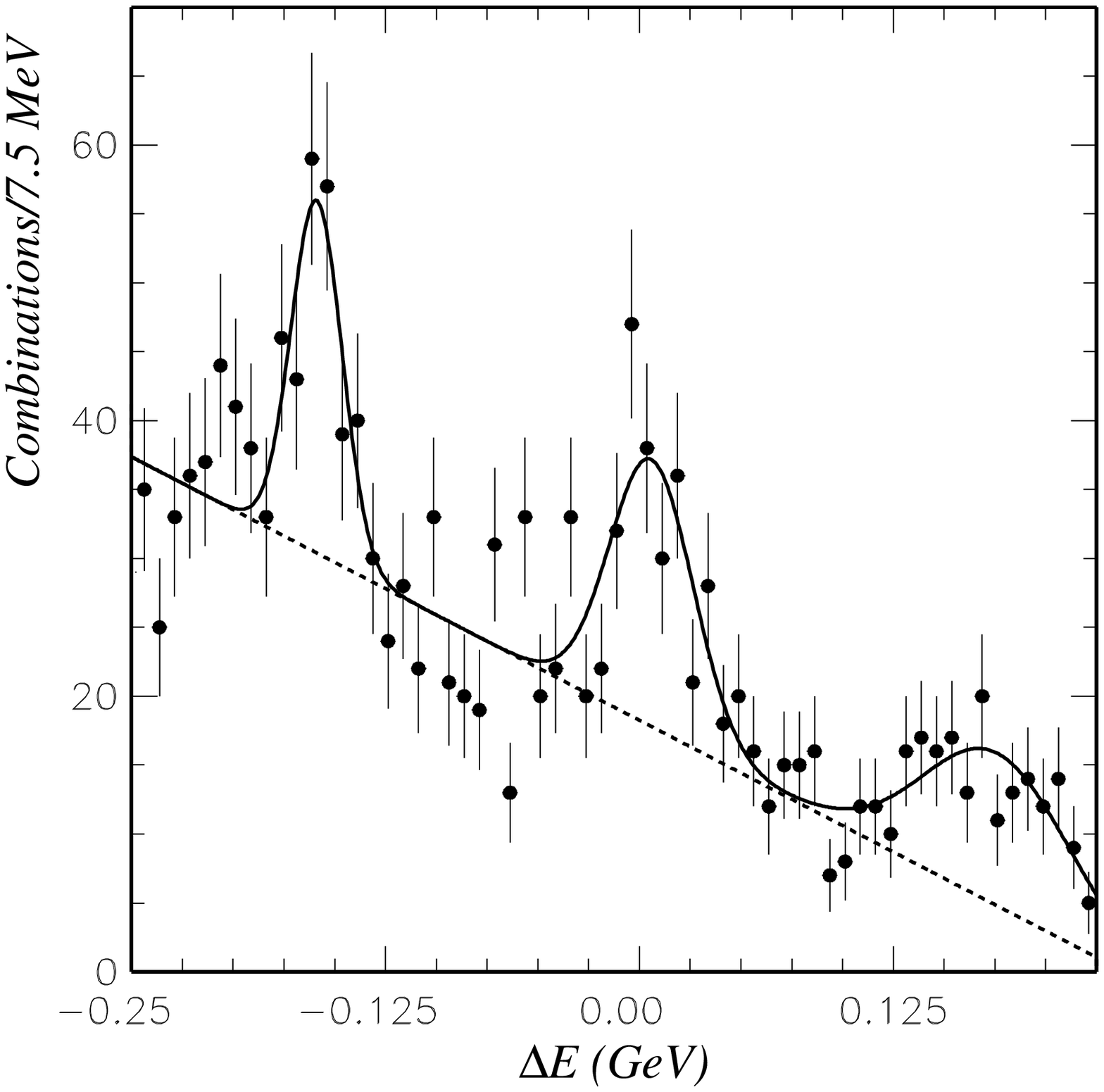}}
\caption{\mes\ and \de\ distributions for the sum of all neutral modes}
\label{Fi:bztoddkall}
\end{center}
\end{figure}

\begin{figure}[H]
\begin{center}
\scalebox{.35}{
\includegraphics{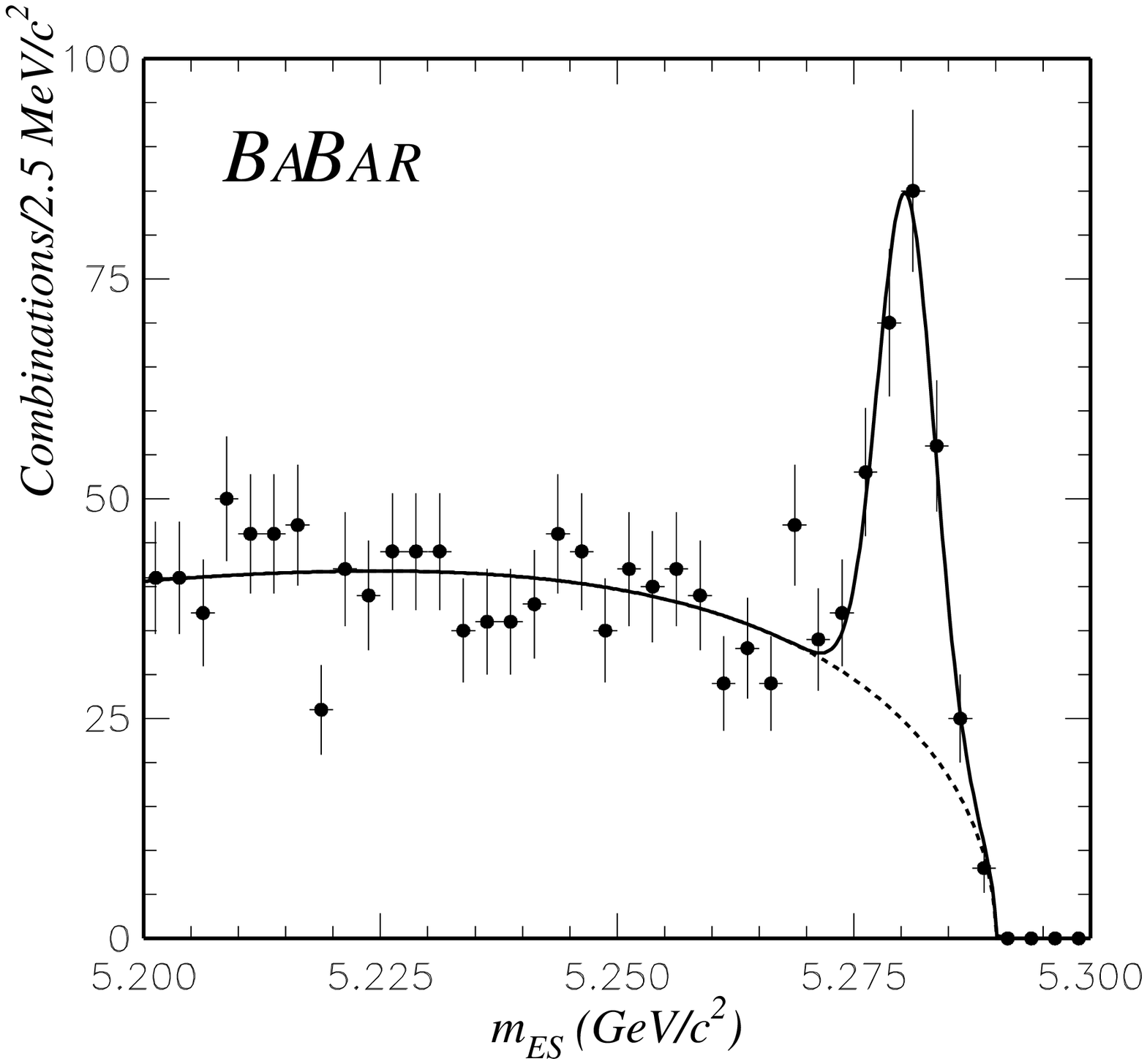}
\includegraphics{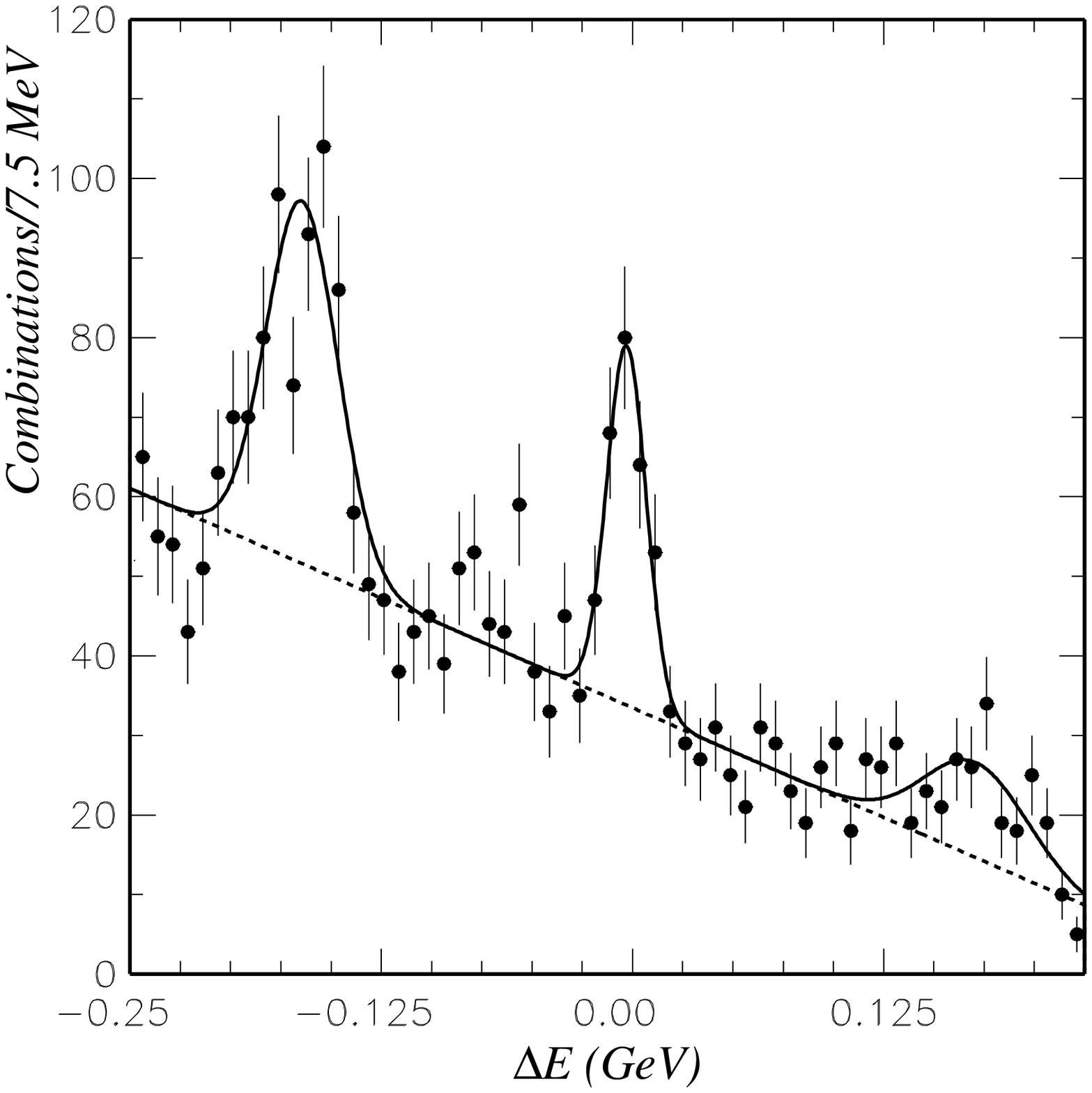}}
\caption{\mes\ and \de\ distributions for the sum of all charged modes}
\label{Fi:bchtoddkall}
\end{center}
\end{figure}

 Since multiple candidates are removed only submode by submode the same
event can appear several times in distributions obtained by summing over all modes. 
In the \de\ distribution this manifests itself as three distinct peaks, as can be seen in
Figs. \ref{Fi:bztoddkall} and \ref{Fi:bchtoddkall}.
An event will appear in the peak near 0\mev\ when reconstructed
correctly, in the peak around $-160$\mev\ if it is a $D^{*}DK$
($D^*D^*K$) decay reconstructed as $DDK$ ($D^*DK$), and near the
peak around $+160$\mev\ if it is a $DDK$ ($D^*DK$) decay reconstructed
as $D^*DK$ ($D^*D^*K$).

About  120 $B^0$'s and 180 $B^\pm$ decays have been  reconstructed. 
The \mes\ distributions (Figs. \ref{Fi:bztoddkall} and  
\ref{Fi:bchtoddkall}) contain only events with $|\de|<24\mev$.
From Monte Carlo studies, the \mes\ resolutions of the different sub-modes are
 quite similar and the \mes\ spectrum of $B^0$ and $B^\pm$ events can be 
 fitted by the sum of a background shape and a Gaussian function used to 
extract the number of signal events. The background is empirically described 
by the function 
$$ { dN \over dm_{ES} } \propto m_{ES} \times 
 \sqrt{1-{m_{ES}^2 \over E_{beam}^{*2}}} \times 
 \exp{\left[-\zeta \left( 1-{m_{ES}^2 \over E_{beam}^{*2}}\right)\right]},$$
 where the only free parameters are $\zeta$ and the normalization factor. 
This function is referred to as the ARGUS function in the following. 
The \de\ distributions (Figs. \ref{Fi:bztoddkall} and \ref{Fi:bchtoddkall})
 contain only events with $\mes > 5.27\mevcc$. They have been fitted by the sum 
of a polynomial  background and three Gaussian functions for the three signal 
components described above. However, the fits to the  \de\ 
distributions are only indicative since they merge many $B$ and $D$ sub-decay modes, 
which have significantly different \de\
resolutions depending on the number of $\pi^0$'s or photons involved 
\section{Measurement of exclusive branching fractions}
In this section, we present measurements of branching fraction for the three decay
channels \bdstdzk, \bdstdstzk\ and \btodsdsk. Several candidates are also observed
in the $CP$ conjugate modes \bztoccz\ but without extracting 
branching fractions.
\subsection{Monte Carlo samples and efficiencies}
The selection efficiencies for each mode were obtained from detailed 
Monte Carlo simulation, in which the detector response is modeled with the 
GEANT3 program \cite{ref:geant}. In addition, data was used whenever possible to 
determine detector performance and the simulation adjusted accordingly. 
$B$ meson decays to $DDK$ were generated with 
a three-body phase space model. For each sub-decay mode, samples of 
5000 signal events were produced. 
Typical efficiencies range from 10\%, for \bdstdzk\ with both $D^0$'s 
decaying to $K\pi$,  to less than 1\%, for \btodsdsk\ with $D^0$'s 
decaying to $K\pi\pi^0$ or $K3\pi$.
\subsection{Systematic uncertainties\label{sec:systDescription}}
Systematic errors account for the uncertainties on tracking and $\pi^0$ 
reconstruction efficiencies, $K$ identification efficiency, $D$ and $B$ 
vertexing requirements, efficiency of the requirement on $\Delta E$ used to 
define the signal box, efficiency of the $D$ mass requirement; uncertainty on 
the background shape; uncertainties on the $D$ and $D^*$ branching 
fractions; uncertainties  on the selection efficiencies arising from
Monte Carlo statistics; and uncertainty on the number of produced $B \overline B$ events
in the data sample. The breakdown of the different contributions to the systematic error 
for each mode is given in Table~\ref{tab:syst_bdstdzk}.
\subsection {\btodsdsk \label{sec:colorsup}}
\begin{figure}[H]
\begin{center}
\includegraphics[height=9cm]{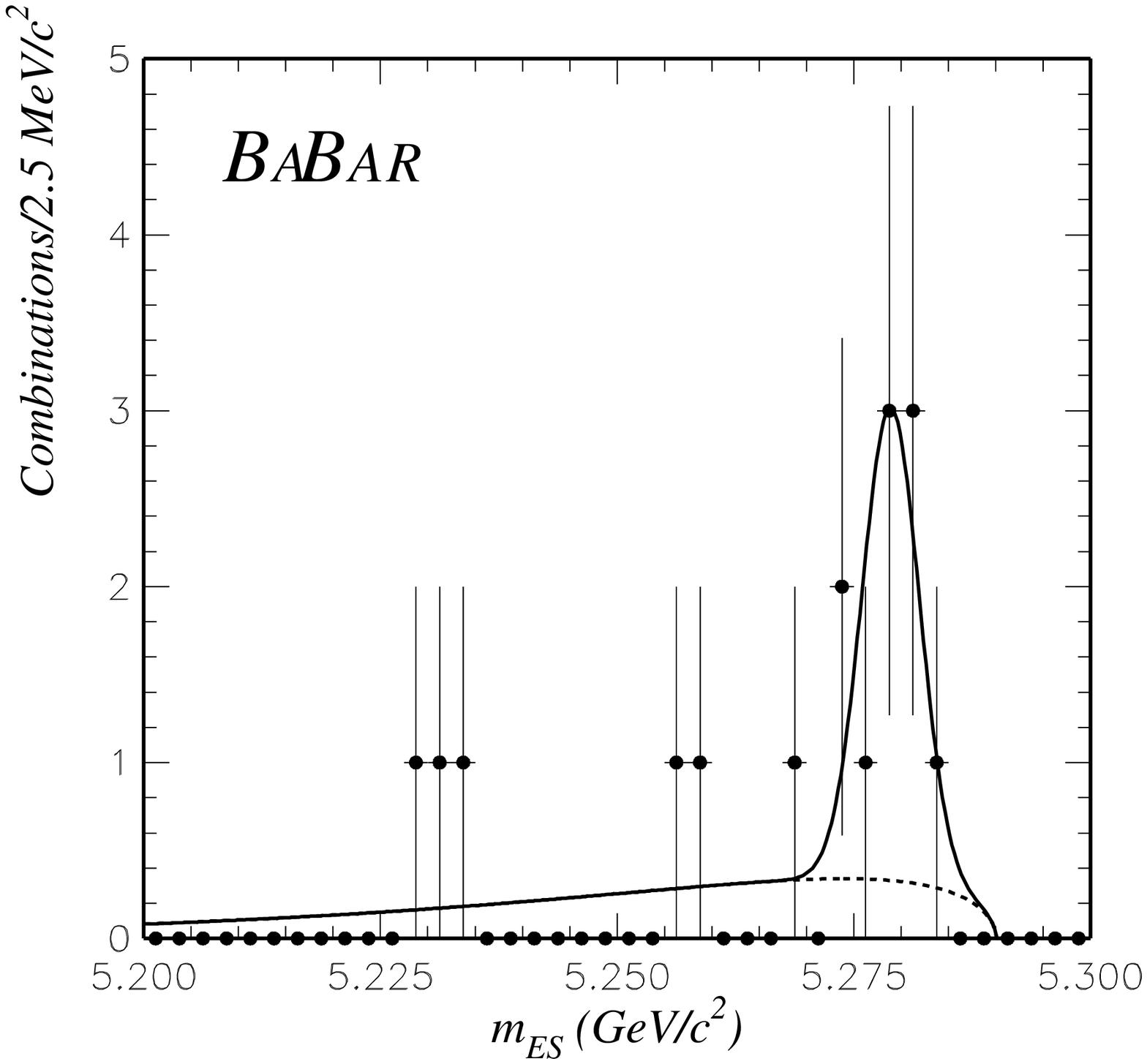}
\caption{\btodsdsk\ \mes\ distribution}
\label{Fi:dsdsk_mes}
\end{center}
\end{figure}

The \mes\ distribution obtained for events with $|\de|<24\mev$ is shown
in Fig. \ref{Fi:dsdsk_mes} for the sum of all the six possible 
$D^0 \times \overline D^0$ decay combinations. A fit to the data is performed 
with the sum of a Gaussian 
function for the signal and an ARGUS function for the background. 
The number of signal events is $8.2\pm 3.5$ and 
the number of background events given by the ARGUS function is 1.7. The 
probability that the signal arises from a background fluctuation is 
$1.4\times 10^{-5}$ ($>5\sigma$). The corresponding preliminary branching fraction is 
measured to be  
$${\cal B}(\btodsdsk) =(3.4\pm 1.6\pm 0.9)\times 10^{-3}$$
The first error quoted is statistical and the second is systematic. The
different contributions to the systematic error are given in Table~\ref{tab:syst_bdstdzk}.
\subsection {$B^0\rightarrow D^{*-}D^{(*)0}K^+$}
\begin{figure}[H] 
\begin{center}
\includegraphics[height=8cm]{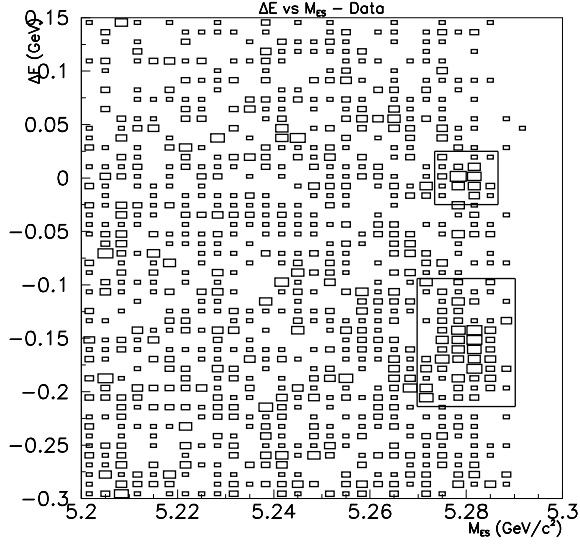}
\end{center}
\caption{Distribution of \de\ versus \mes\ for $D^{*-}D^0K^+$ combinations 
in the data. The signal boxes are defined by a $\pm 3 \sigma$ requirement on \mes. 
The box $|\de|< 25 \mev$ corresponds to \bdstdzk\ decays, while the box 
$|\de + 154 |< 60 \mev$ corresponds dominantly to decays \bdstdstzk\ 
(The $\pi^0$ or $\gamma$ from $D^{*0}\rightarrow D^0\pi^0$ or $D^0\gamma$ is not 
reconstructed here)} 
\label{fig:dembdata_V4}
\end{figure}

In this analysis we require that either the  $D^0$ or the $\overline D^0$ 
decays to $K\pi$ and we do not explicitly reconstruct the $\pi^0$ or 
the photon from $D^{*0}\rightarrow D^0\pi^0$ or $D^0\gamma$.
The \mes\ versus \de\ distribution of $D^{*-}D^0K^+$ 
combinations is shown in Fig. \ref{fig:dembdata_V4} for the sum of the 
three $D^0 \overline D^0$ sub-modes considered. 
Despite the background level, two separate enhancements, due to
the decay modes  \bdstdzk\ and \bdstdstzk, are clearly visible. The 
enhancement in the region $\de \simeq 0$, $\mes \simeq 5.28$\gevcc 
corresponds to decays \bdstdzk, while the second enhancement in the region 
$\de \simeq -154$\mev, $\mes \simeq 5.28$\gevcc corresponds to decays 
\bdstdstzk. 

\begin{figure}[H] 
\begin{center}
\includegraphics[height=8cm]{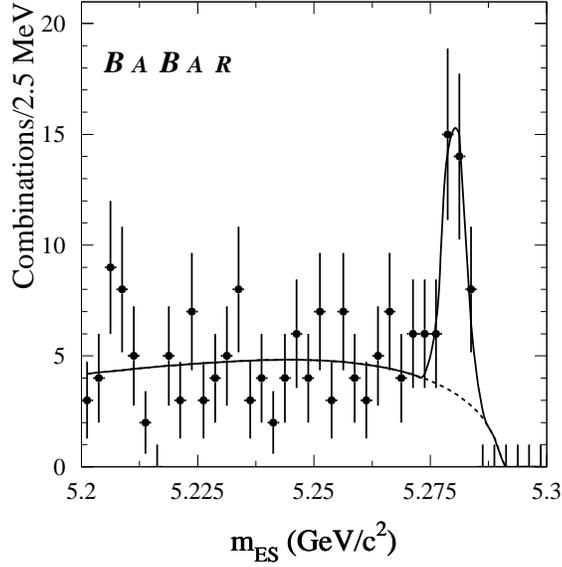}
\end{center}
\caption{Distribution of \mes\  for $D^{*-}D^0K^+$ combinations with 
$|\de|< 25 \mev$. An ARGUS background function is used together with 
a Gaussian for the signal shape to fit the data.}
\label{fig:mbddkdatafit_V4}
\end{figure}

Events containing \bdstdzk\ decays are selected by requiring $|\de|<25\mev$. 
The \mes\ spectrum for these events is shown in  
Fig. \ref{fig:mbddkdatafit_V4} along with a fit with the sum 
of a Gaussian function describing the signal and an ARGUS function describing 
the background. The number of signal events is found 
to be $29.6 \pm 7.2 $. After correcting for the selection efficiencies 
and for the intermediate $D^0$ and $D^{*+}$  branching fractions 
\cite{ref:pdg}, the preliminary branching fraction for \bdstdzk\ is found to be  
$$
{\cal B}(B^0 \ra D^{*-}D^{0}K^+) = (2.8 \pm 0.7 \pm 0.5)\times 10^{-3}, 
$$
where the first  error quoted is statistical and the second systematic. 
The breakdown of the various contributions to the systematic error 
is given in Table~\ref{tab:syst_bdstdzk}. 

\begin{figure}[H] 
\begin{center}
\includegraphics[height=8cm]{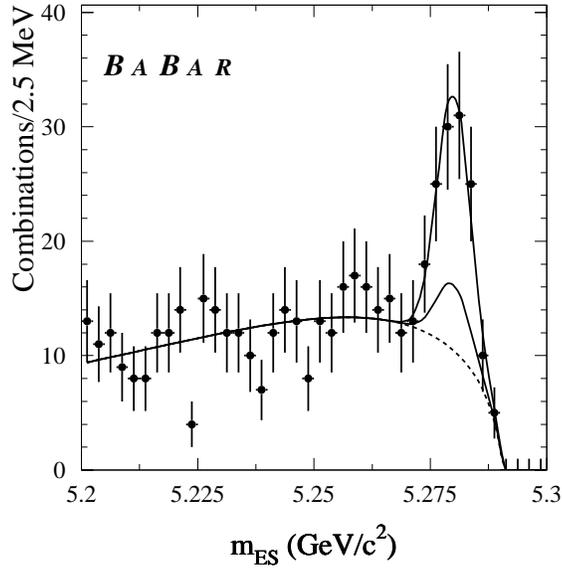}
\end{center}
\caption{Distribution of \mes\ for $D^{*-}D^0K^+$ combinations with 
$|\de+154|<60\mev$ ($B\rightarrow D^{*-}D^{*0}K^+$ signal region). 
An ARGUS background function is used together with 
a Gaussian for the signal shape to fit the data. The 
\btodsdsk\ background contribution is shown as a small Gaussian on top of the 
combinatorial background shape.}
\label{fig:mbddstkdatafit_V4}
\end{figure}

Events containing \bdstdstzk\ decays are selected by requiring $|\de+154|<60 \mev$.
The average position and width of \de\ for \bdstdstzk\ is found to be in good 
agreement with expectations from \bdstdstzk\ signal Monte Carlo studies. 
The \mes\ spectrum of the selected events is shown in 
Fig. \ref{fig:mbddstkdatafit_V4} along with a fit with the sum of 
a Gaussian and an ARGUS background function. The number of signal 
events found is $80.2 \pm 15.3 $.
\par To extract the \bdstdstzk\ branching fraction, the contamination from 
decays $\btodsdsk$, where the $\pi^+$ from the $D^{*+}$ is not reconstructed, 
needs to be subtracted. This contribution has been estimated by performing the 
 \bdstdstzk analysis on $\btodsdsk$ signal Monte Carlo, assuming 
the $\btodsdsk$ branching fraction presented in Section~\ref{sec:colorsup}.
The $\btodsdsk$ background contribution is shown in 
Fig.~\ref{fig:mbddstkdatafit_V4} as a small Gaussian on top of the 
combinatorial background shape; it is estimated to be $20.6\pm 9.7$ events.
After subtracting this contribution, the preliminary \bdstdstzk\ branching fraction is 
determined to be: 
$$
{\cal B}(B^0 \ra D^{*-}D^{*0}K^+) = (6.8 \pm 1.7 \pm 1.7)\times 10^{-3} 
$$
where the last uncertainty is systematic. The breakdown of the various 
contributions to the systematic error is given in Table~\ref{tab:syst_bdstdzk}. 

\begin{table}[H]
\begin{center}
\caption{ Breakdown of the various contributions to the relative systematic 
uncertainty on the \btodsdsk, \bdstdzk\ and  \bdstdstzk\ branching fraction measurements.}
\vskip 0.2cm
\label{tab:syst_bdstdzk} 
\begin{tabular}{|c|c|c|c|}
\hline
  	               & \btodsdsk & \bdstdzk\ & \bdstdstzk \\
  Source               & error(\%) & error(\%) & error(\%) \\ \hline\hline
Tracking + 
Neutral efficiency 	& 9.7 & 8.8 	& 8.8 \\
Vertexing efficiency	& 10  & 5.6     & 8.3 \\
PID efficiency	        & 9   & 5.3 	& 5.3 \\
\de\ requirements	& 2   & 7.7 	& 2.4 \\
$D$ meson mass requirements      & 13.4&  -      &  -  \\
Intermediate BF		& 5.6 & 5.6  	& 7.5 \\
Background shape 	&  -  & 4.9     & 2.9 \\
Monte Carlo statistics  & 16  & 3.5 	& 4.3 \\
N$_{B \overline B}$		& 1.6 & 1.6 	& 1.6 \\
$\btodsdsk\ $ bkg		& -   & -       & 19.4\\ \hline\hline

{\bf Total	} 	& {\bf 27}  & {\bf 16.3} 	& {\bf 25.4}\\
\hline
\end{tabular}
\end{center}
\end{table}

\subsection {\bztoccz}

\begin{figure}[H]
\begin{center}
\includegraphics[height=8cm]{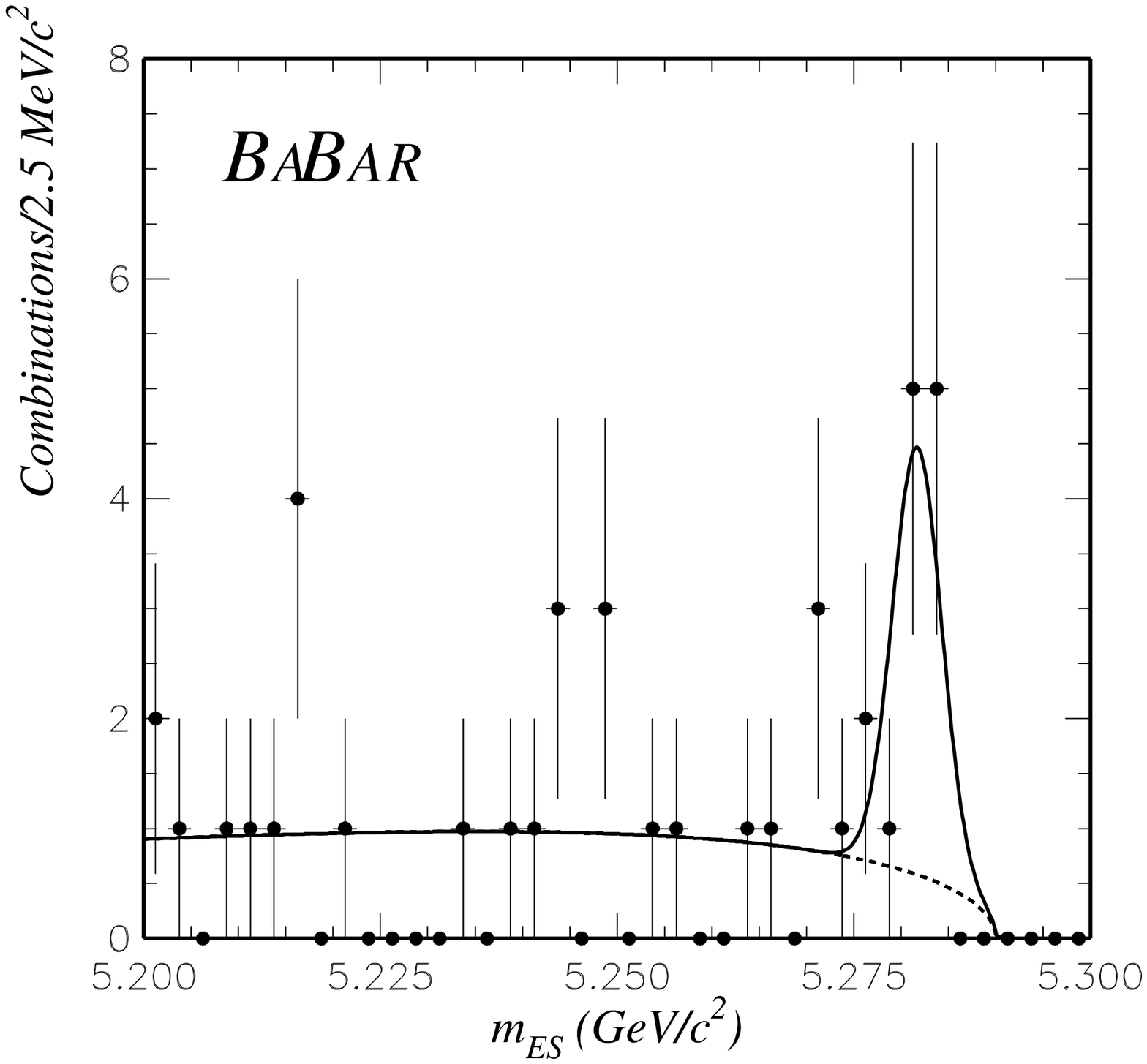}
\caption{\bztoccz\ \mes\ distribution}
\label{Fi:cpmes}
\end{center}
\end{figure}

The \mes\ distribution for events reconstructed in the channels
\bztoccz\ is shown in Fig.~\ref{Fi:cpmes}. For modes involving $D^0$'s, at 
least one decay $D^0\rightarrow K\pi$ was required. The fitted number of 
signal events is $10.1\pm 3.7$ with an estimated background of 3.4 events. 
The probability that the signal is a fluctuation
of the background is $1.4\times 10^{-5}$ ($>5\sigma$). Most of the signal is due 
to the channels $B^0\rightarrow D^{*+}D^-K^0_S$ ($4.7\pm 2.2$ events 
with a background of 1 event) and $B^0\rightarrow D^{*+}D^{*-}K^0_S$ ($4.8\pm 2.2$ events 
with a background of 0.3 event). As pointed out in \cite{ref:browder3}, 
the channel $B^0\rightarrow D^{*+}D^{*-}K^0_S$ is a \CP\ conjugate state that could 
be used for  $\sin 2\beta$ measurements. However, given the presently observed rate for
reconstructing events, large improvements in the selection efficiencies are still 
needed before challenging the ``golden'' channels $B^0\rightarrow D^{*+}D^{*-}$
as suggested in \cite{ref:browder3}.  


\section{Summary}
\label{sec:Summary}
Using about 23M $B\overline B$ events, we have observed several hundred completely 
reconstructed $B\rightarrow D^{(*)}\overline D^{(*)}K$ decays. The following preliminary branching 
fractions have been measured:  
$$ {\cal B}(B^0 \ra D^{*-}D^{0}K^+) = (2.8 \pm 0.7 \pm 0.5)\times 10^{-3} $$
$$ {\cal B}(B^0 \ra D^{*-}D^{*0}K^+) = (6.8 \pm 1.7 \pm 1.7)\times 10^{-3} $$
in good agreement with the CLEO measurements ${\cal B}(B^0 \ra D^{*-}D^{0}K^+) 
= (4.5^{+2.5}_{-1.9} \pm 0.8)\times 10^{-3}$ and $ {\cal B}(B^0 \ra D^{*-}D^{*0}K^+) 
= (13.0^{+7.8}_{-5.8} \pm 3.6 )\times 10^{-3} $ \cite{ref:cleoddk}. 

We have observed an excess of $8.2\pm 3.5$ events over a background of 1.7 events in 
the color-suppressed decay mode $B^+\rightarrow D^{*+}D^{*-}K^+$, where
no significant number of candidates has been previously seen.
The corresponding 
preliminary branching fraction is measured to be 
$$ {\cal B}(\btodsdsk)= (3.4\pm 1.6\pm 0.9) \times 10^{-3} $$ 
Finally, several candidates have also been observed in the \CP\ conjugate states 
$B^0\rightarrow D^{(*)+}D^{(*)-}K^0_S$.
This study confirms that the transitions  $b \rightarrow c \overline c s$ can 
proceed through the decays  $B\rightarrow D^{(*)}\overline D^{(*)}K$. To quantify 
more precisely this statement, we intend to measure all the individual 
$B\rightarrow D^{(*)}\overline D^{(*)}K$ branching fractions and study the decay 
kinematics of these decays in the near future.

\section{Acknowledgments}
\label{sec:Acknowledgments}
We are grateful for the 
extraordinary contributions of our \pep2\ colleagues in
achieving the excellent luminosity and machine conditions
that have made this work possible.
The collaborating institutions wish to thank 
SLAC for its support and the kind hospitality extended to them. 
This work is supported by the
US Department of Energy
and National Science Foundation, the
Natural Sciences and Engineering Research Council (Canada),
Institute of High Energy Physics (China), the
Commissariat \`a l'Energie Atomique and
Institut National de Physique Nucl\'eaire et de Physique des Particules
(France), the
Bundesministerium f\"ur Bildung und Forschung
(Germany), the
Istituto Nazionale di Fisica Nucleare (Italy),
the Research Council of Norway, the
Ministry of Science and Technology of the Russian Federation, and the
Particle Physics and Astronomy Research Council (United Kingdom). 
Individuals have received support from the Swiss 
National Science Foundation, the A. P. Sloan Foundation, 
the Research Corporation,
and the Alexander von Humboldt Foundation.


\begin{thebibliography}{99}


\bibitem{ref:dsdargus} ARGUS Collaboration, H.~Albrecht \emph{et al.}, 
Z.~Phys. {\bf C54}, 1 (1992).

\bibitem{ref:dsdcleo} CLEO Collaboration, D.~Gibaut \emph{et al.},  
Phys.~Rev. {\bf D53}, 4734 (1996).

\bibitem{ref:dsdcleo2} CLEO Collaboration, S.~Ahmed \emph{et al.}, 
Phys.~Rev. {\bf D62}, 112003 (2000)

\bibitem{ref:dsdbabar} \babar\ Collaboration, B.\ Aubert {\em et al.}, 
 \babar-CONF-01/27.

\bibitem{ref:browder2}  
 T.~Browder, \emph{Hadronic decays and lifetimes of $B$ and D mesons}, 
 Proceedings of the 1996 Warsaw ICHEP conference, 
Z.~Ajduk and A.K.~Wroblewski Eds, World Scientific (1997) p1139.

\bibitem{ref:bigi} I.~I.~Bigi, B.~Blok, M.~Shifman and A.~Vainshtein, 
Phys.~Lett. {\bf B323}, 408 (1994).

\bibitem{ref:buchalla} G.~Buchalla, I.~Dunietz and H.~Yamamoto, 
Phys.~Lett. {\bf B364}, 188 (1995). 

\bibitem{ref:cleoupv} CLEO Collaboration,
T.E.Coan et al.,  
Phys.~Rev.~Lett. {\bf 80}, 1150 (1998)

\bibitem{ref:cleoddk} CLEO Collaboration, CLEO~CONF~97-26, EPS97~337.

\bibitem{ref:alephddk} ALEPH Collaboration, R.~Barate \emph{et al.}, 
Eur.~Phys.~J.{\bf C4}, 387-407 (1998).

\bibitem{ref:babar}
\babar\ Collaboration, B.\ Aubert {\em et al.}, SLAC-PUB-8596, hep-ex-0105044,
 to be published in Nucl.Inst.Methods. 

\bibitem{ref:geant} ``GEANT, Detector Description and Simulation Tool'',
CERN program library longwritup W5013(1994)

\bibitem{ref:pdg}
Particle Data Group,  D.E.~Groom \emph{et al.},  
Eur.~Phys.~J.~{\bf C15}, 1 (2000)

\bibitem{ref:browder3}
T.~Browder \emph{et al.}, 
Phys.~Rev~{\bf D61}, 054009 (2000)

\end{thebibliography}
\end{document}